\documentclass[11pt]{scrartcl}
\usepackage{amsmath}
\usepackage{amssymb}
\usepackage{amsthm}
\usepackage{slashed}
\usepackage{graphicx}
\usepackage{color}
\usepackage[centering,headsep=15pt,head=80pt,foot=20pt,margin=60pt]{geometry}
\usepackage[scaled]{helvet}
\usepackage[slantedGreek]{mathptmx}

\newcommand{\Group}[2]{{ \hbox{{\itshape{#1}}($#2$)} }}
\newcommand{\U}[1]{\Group{U\kern0.05em}{#1}}
\newcommand{\SU}[1]{\Group{SU\kern0.1em}{#1}}
\newcommand{\SL}[1]{\Group{SL\kern0.05em}{#1}}
\newcommand{\Sp}[1]{\Group{Sp\kern0.05em}{#1}}
\newcommand{\SO}[1]{\Group{SO\kern0.1em}{#1}}

\newcommand{\mybar}[1]%
    {{\kern 0.8pt\overline{\kern -0.8pt#1\kern -0.8pt}\kern 0.8pt}}
\newcommand{\roughly}[1]%
    {{ \mathrel{\raise.3ex\hbox{ $#1$\kern-.75em\lower1ex\hbox{$\sim$}} } }}

\newcommand{\avg}[1]{\langle #1 \rangle}

\newcommand{\nop}[1]{:\kern-.3em#1\kern-.3em:}

\newcommand{\al}{\ensuremath{\alpha}}

\newcommand{\de}{\ensuremath{\delta}}
\newcommand{\De}{\ensuremath{\Delta}}

\newcommand{\et}{\ensuremath{\eta}}

\newcommand{\si}{\ensuremath{\sigma}}

\newcommand{\ch}{\ensuremath{\chi}}

\newcommand{\GeV}{ \ensuremath{\mathrm{~GeV}} }

\numberwithin{equation}{section}
\numberwithin{figure}{section}

\begin{document}
\begin{titlepage}
\begin{flushright}
MHDP
\end{flushright}
\vskip 5em
\begin{center}
{\Large \bfseries
 The dual King relation\\
}
\vskip 4em
\renewcommand{\thefootnote}{\alph{footnote}}

Yasuhiro Yamamoto$^\sharp$

\vskip 4em
$^\sharp$
	\textit{Physics Division, National Center for Theoretical Sciences, National Taiwan University, Taipei 10617, Taiwan}
\vskip 4em
\textbf{Abstract}
\end{center}
\medskip
\noindent

We introduce a new linear relation in the isotope shifts of atomic spectroscopy.
While the famous King relation is the linear relation among the different transitions, the new one is the linear relation among the different isotope pairs.
Since we obtain this relation by exchanging the roles of the electronic and the nuclear factors in the original relation, we call it the dual King relation.
This relation shows us similar information to the original King relation without including new fit parameters when we measure the isotope shifts of many transitions.
In the dual King relation, the fit coefficients consist of the nuclear factors.
Then, the fit results give us constraints to the ratios of the isotope dependence independent of the electron wave functions.
This helps us to test the origin of unknown higher order isotope shifts.
We show that the dual King relation can also be employed to constrain the weakly interacting light new boson at the same level as the original King relation.

\bigskip
\vfill
\end{titlepage}
\tableofcontents
\setcounter{footnote}{0}
\section{Introduction}

The weakly interacting light boson is an attractive candidate included in the model beyond the Standard Model.
Such a new force career is introduced in the new sector to explain the remaining mysteries in the Standard Model, while the interactions with the Standard Model particles are kept weak.

The light boson itself can play a role in solving the observed problems in astrophysical observations, e.g., the dark matter distribution~\cite{1705.02358}, the anomalous stellar cooling~\cite{1512.08108}, etc., and also low-energy experiments like the muon anomalous magnetic moment~\cite{2006.04822}.
While the interaction with the Standard Model particle should be small, the particle can appear in other low energy experiments~\cite{1504.04855,1702.03327} and cosmological observations~\cite{2112.11096}.
Hence, the weakly interacting light boson attracts the attention of physicists in different fields.

The isotope shift of the atomic spectrum is one of the phenomena where the effect of the new force can appear~\cite{1601.05087,1704.05068}.
The technology of precision atomic spectroscopy is continuously improving and the precision of the isotope shift measurements reach at the 10 mHz level with the state of the art optical lattice technology~\cite{RefTakano,Manovitz:2019czu}.
The isotope shift is traditionally used to determine the isotope dependence of the nuclear mean square charge radii~\cite{Angeli:2013epw,1602.07906}.
In these days, the constraint on the weakly interacting light boson is also actively studied with the isotope shift, and then it is rapidly improved.

The method to obtain the constraint is based on the King relation and its generalization, which is the linear relation among the isotope shifts for the distinct isotope pairs~\cite{RefKing,1710.11443}.
If electrons and neutrons couple to the new force, the effect appears as the violation of the relation.
If the violation is observed, it can be attributed to not only the new force but also the higher order isotope shifts~\cite{RefBlundell,1709.00600,1710.11443,1911.05345}.
The higher order contributions can be suppressed by the generalized King relation as long as we have enough pairs of the isotopes to eliminate them from the relation.
Therefore, the number of distinct isotope pairs is important to analyse the precision isotope shift data while the stable isotope is limited by nature.
Compared to isotopes, the number of transitions is less limited.
We already have five precision isotope shifts with Ytterbium~\cite{2004.11383,2110.13544,Figueroa:2022mtm,2201.03578}.
For Calcium, it is also planned to employ highly charged ions~\cite{2102.02309}.
However, including more transitions, we need more parameters to fit the data, and then, the analysis becomes more complicated~\cite{2201.03578,RefNew}.

In this paper, we introduce a new linear relation among the isotope shifts.
The relation is obtained by exchanging the roles of the electronic and the nuclear factors in the original King relation, so we call it the dual King relation.
Each isotope pair is plotted in the space spanned by the different transitions in the original King relation.
In the dual King relation, each transition is plotted in the space of the isotope pairs.
Then, an additional transition follows the same linear relation without including a new parameter.

In the rest of this paper, we formulate the dual King relation in Sec.~\ref{SecKing}.
We show that the fit coefficients in this relation consist of the nuclear isotope dependences.
Then, with the precision measurements of the isotope shifts, the isotope dependences of the nucleus are restricted.
In Sec.~\ref{SecResult}, we analyze the current data of Yb and fake data of Ca with the dual King relation.
We show that, in terms of the linear fit, the dual King relation gives the same results as the original King relation, but with less parameters
In addition, with the fake data of Ca, we demonstrate how the isotope shifts constrain the nuclear isotope dependences independent of the electron wave function.
It helps us to test the origin of unknown higher order isotope shifts.
Here, we also discuss the expected upper bounds for the weakly interacting light new boson.
Section~\ref{SecCon} is the conclusion.

\section{The formulation}
\label{SecKing}

We introduce the isotope shift of an isotope pair $a$ for any transition by the following form,
\begin{align}
 \de\nu_a = K \de\mu_a +F \de\avg{r^2}_a +\sum_i I_i \de\et_{ia},
\label{EqIs}
\end{align}
where $\de\mu$, $\de\avg{r^2}$ and $\de\et_i$ stand for isotope dependences, and the other isotope independent factors $K$, $F$ and $I_i$ are given by electron wave functions for each transition. 
The index of the transition is suppressed.
We obtain the dual King relation with the set of $n+1$ isotope shifts as
\begin{align}
  \de\vec{\nu}
 =&
  \De \cdot \vec{S} \\
 =&
  \begin{pmatrix}
    \de\mu_1 & \de\avg{r^2}_1 & \cdots & \de\et_1 & 0 \\
    \vdots      & \vdots            & \ddots & \vdots       & 0 \\
    \de\mu_n & \de\avg{r^2}_n & \cdots & \de\et_n & 0 \\
    \de\mu_{n+1} & \de\avg{r^2}_{n+1} & \cdots & \de\et_{n+1} & -1 
	\end{pmatrix}
	\begin{pmatrix}
    K \\ F \\ \vdots \\ I \\ \de\nu_{n+1} 
	\end{pmatrix},
\label{EqKingmat}
\end{align}
where $\de\vec{\nu} = (\de\nu_1, \cdots, \de\nu_n, 0)$.
The matrix $\De$ can be formally inverted as
\begin{align}
   \De^{-1}
 =&
   \begin{pmatrix} \De_{nn} & 0 \\ \De_{n+1} & -1 \end{pmatrix}^{-1} \\
 =&
   \begin{pmatrix} (\De_{nn})^{-1} & 0 \\ \De_{n+1} (\De_{nn})^{-1} &-1 \end{pmatrix}.
\end{align}
If we have additional isotope pairs not to be employed to eliminate higher order contributions, they are easily included in the bottom of Eq.~\eqref{EqKingmat} as additional linear relations\footnotemark.
\footnotetext{
 A similar linear relation is mentioned as the mass independent linearity in Ref.~\cite{2005.06144}.
 However, it is clearly a part of the generalized King relation introduced in Ref.~\cite{1710.11443} if one can use the linear algebra.
}
The original King relation is interpreted as the equations to obtain the nuclear factors by the measured isotope shifts and the electronic factors.
The dual King relation shows the equations to obtain the electronic factors by the isotope shifts and the nuclear factors.

We need at least three isotope pairs for the new relation, which is, using the coefficients $\al_{1,2}$, written as
\begin{align}
  \de\nu_3 =& \al_1 \de\nu_1 + \al_2 \de\nu_2.
\label{EqDking}
\end{align}
This means that the isotope dependence of the leading isotope shifts satisfies
\begin{align}
  \de\avg{r^2}_3 =& \al_1 \de\avg{r^2}_1 +\al_2 \de\avg{r^2}_2 ,\\
  \de\mu_3 =& \al_1 \de\mu_1 +\al_2 \de\mu_2,
\end{align}
and then, the coefficients can be solved as
\begin{align}
  \al_1 =& \frac{\de\mu_3 \de\avg{r^2}_2 -\de\mu_2\de\avg{r^2}_3}{\de\mu_1 \de\avg{r^2}_2 -\de\mu_2 \de\avg{r^2}_1}, \\
  \al_2 =& \frac{\de\mu_3 \de\avg{r^2}_1 -\de\mu_1\de\avg{r^2}_3}{\de\mu_2 \de\avg{r^2}_1 -\de\mu_1 \de\avg{r^2}_2}.
\label{EqDualco}
\end{align}
It shows us that the ratios of the isotope dependences $\de\mu$ and $\de\avg{r^2}$ are directly constrained by atomic spectroscopy as the coefficients of the linear relation.

At a level of precision, a new higher order isotope shift violates the above linear relation.
If it is a possible known higher order contribution, it can be subtracted as it is done in the generalized King relation.
We write such a higher order term as $I \de\et_a$, then the relation is modified as
\begin{align}
  \de\nu_3 =& \al_1 \de\nu_1 + \al_2 \de\nu_2 +I (\de\et_3 -\al_1 \de\et_1 -\al_2 \de\et_2).
\label{EqDualplus}
\end{align}
To fit this relation with the additional term, we need to know each electronic factor $I$ with sufficient precision.
This is equivalent to employing the additional linearity to be solved for $I$ in the higher dimensional linear relation when we have another isotope pair.

Higher order isotope shifts do not always violate the linear relation, even if the data include the contributions.
According to the estimate in Refs.~\cite{RefBlundell,1709.00600,1911.05345}, the sizes of higher order contributions are $O(10^{-4})$ to the leading contributions.
One of the candidates $\de\avg{r^4}$ discussed in Refs.~\cite{1911.05345,2201.03578} is approximately proportional to $\de\avg{r^2}$ for the first a few digit, see Ref.~\cite{2004.11383}.
Such a contribution can be involved in the leading contributions by the constant shift of the electronic or the nuclear factors keeping the linear relation.

The constant shifts for the nuclear factors are independent of the transitions but can depend on the isotope pairs.
Assuming that they are also independent of the pairs, we can shift them by factors $a_m$ and $a_r$ as
\begin{align}
  \de\mu_a & \rightarrow \de\mu_a +a_m ,\\
	\de\avg{r^2}_a & \rightarrow \de\avg{r^2}_a +a_r. 
\end{align}
They are solved as
\begin{align}
 a_m =& -\frac{\de\mu_3 -\al_1 \de\mu_1 -\al_2 \de\mu_2}{1-\al_1 -\al_2}, \\
 a_r =& -\frac{\de\avg{r^2}_3 -\al_1 \de\avg{r^2}_1 -\al_2 \de\avg{r^2}_2}{1-\al_1 -\al_2}.
\end{align}

For example, a heavy new boson generate such a shift.
According to Ref.~\cite{1710.11443}, its leading order contribution is given by $a_m=0$ and $a_r= (-1)^{s+1} 3 g_e g_n /(Z\pi \al m^2)$ with $\de A=2$, where $g_{e,n}$ are the coupling constants of electron and neutron with the new light boson.
If Calcium is employed to constrain the heavy particle shift, we obtain $m^2/(g_e g_n) \gtrsim 5\times 10^{-4}/x \GeV^2$.
Here, we assume that the relative errors of $\de\avg{r^2}$ are given by $x$ independent of the isotope pairs.
This numerical coefficient in the formula is of the same order with other atoms.
For example, to reach the vector coupling of $Z$ boson, we need the relative error of the mean square charge radius $x\sim 10^{-10}$.
Even if this simple calculation does not work well, we still need the precision of $\de\avg{r^2}$ measurements at the level of $a_r/\de\avg{r^2}\sim 10^{-7}$ to observe this contribution of the leading order.

\section{Numerical analysis}
\label{SecResult}

The recent precision data of the Ytterbium isotope shifts are given by Refs.~\cite{2004.11383,2110.13544,Figueroa:2022mtm,2201.03578}.
Since there are five viable isotopes for Yb, we can cancel an additional source of the isotope shift.
However, as shown in Refs.~\cite{2110.13544,2201.03578}, the data includes two additional sources. 
Hence, the measured data are not linearly aligned in the 3D dual King relation.
The 2D dual King fit of the Yb data gives us the minimum $\ch^2$ of $1.7\times 10^4$.
The 3D result is $\ch^2 =26$ with 9 degrees of freedom, which corresponds to $3.1\si$ deviation.
By the transitive consistency condition, this minimum $\ch^2$ is larger than 9.1.
This result shows us the non linearity is inevitable in the known precise Yb isotope shift data.
For the original King relation, the corresponding analysis is provided in Ref.~\cite{RefNew}.

To demonstrate the analysis with many transitions, we employ the isotope shifts with the highly charged calcium ions discussed in Ref.~\cite{2102.02309}.
We employ a part of their numerical results of the electronic factors.
To keep consistency with the known measurements, we also employ the electronic factors given by Ref.~\cite{Kramida:2020noe}.
The fake data are generated with these electronic factors and the nuclear masses calculated with Ref.~\cite{Huang:2021nwk,Wang:2021xhn} by the original 2D King relations.
The data are consistent with the observed isotope shifts about 10 \%.
This means that the exploit values in the following analyses are away from the real values.
However, their relative errors show us some expectations of the results we obtain from the future measurements.
Therefore, we shift the central values of the following fit to coincide with the experimental values.

Here, we employ four isotope pairs and seven transitions.
To fit these data with the original King relation, we need six 2D or five 3D linear relations, namely, 12 or 15 fit parameters.
On the other hand, the 2D and the 3D dual King relations require four and three parameters, respectively.

The generated data are further modified to be consistent with the 3D original King relation.
Then, we obtained $\ch^2 = 3.5$ with the 2D dual King relation of an error of 1 Hz to embed a higher order contribution at this level, as expected by Refs.~\cite{1709.00600,1911.05345}.
This $\ch^2$ scales as proportional to the inverse square of the error.
Then, the embedded non-linearity is suppressed by the 3D dual King relation, where $\ch^2=7.4\times 10^{-3}$ with an error of 10 mHz. 
Therefore, the fake data behaves as if an unknown higher-order isotope shift appears at the 1 Hz level.

\begin{figure}[hbt]
\centering
\includegraphics[scale=0.6,clip]{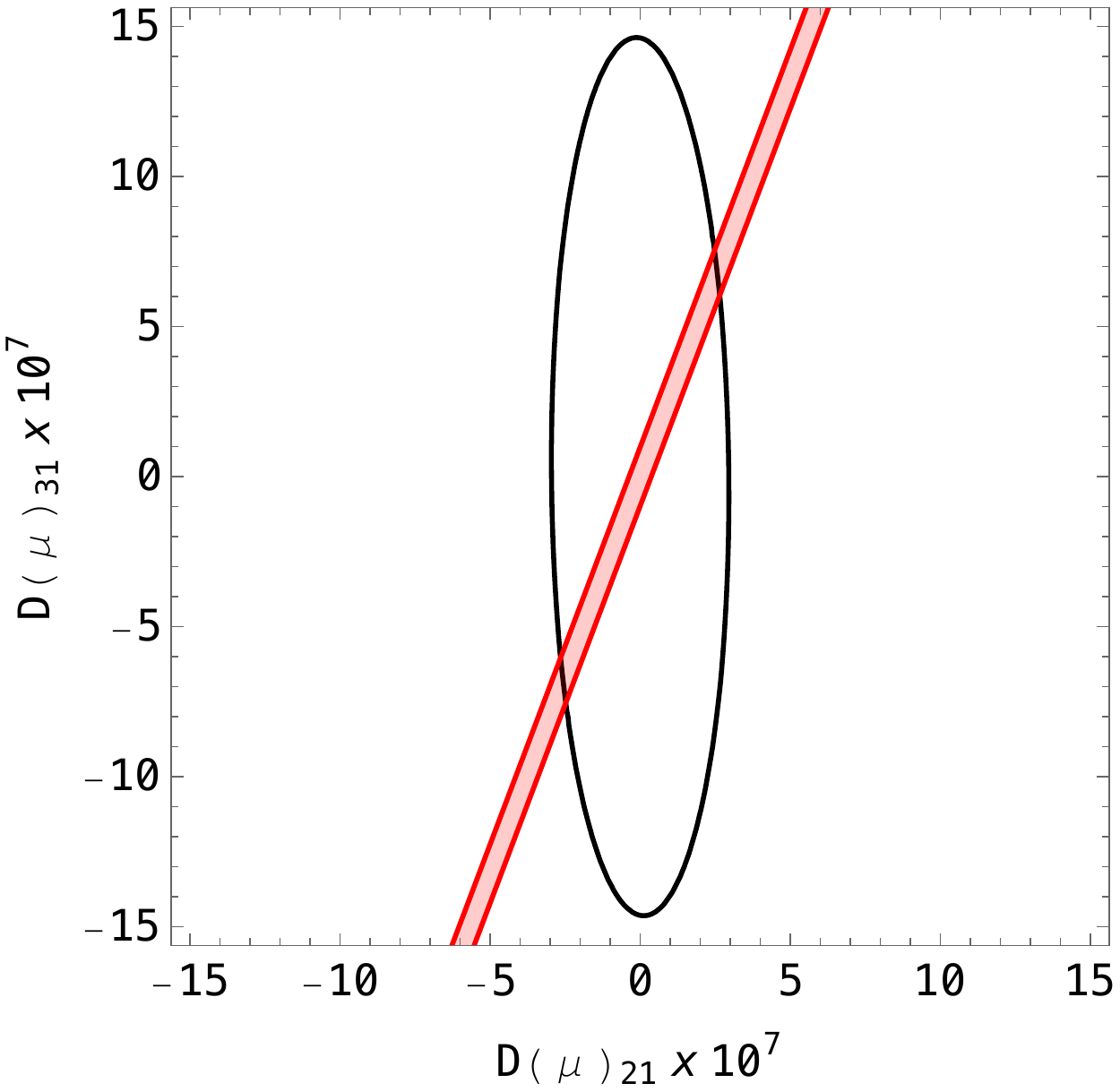}
\phantom{MM}
\includegraphics[scale=0.65,clip]{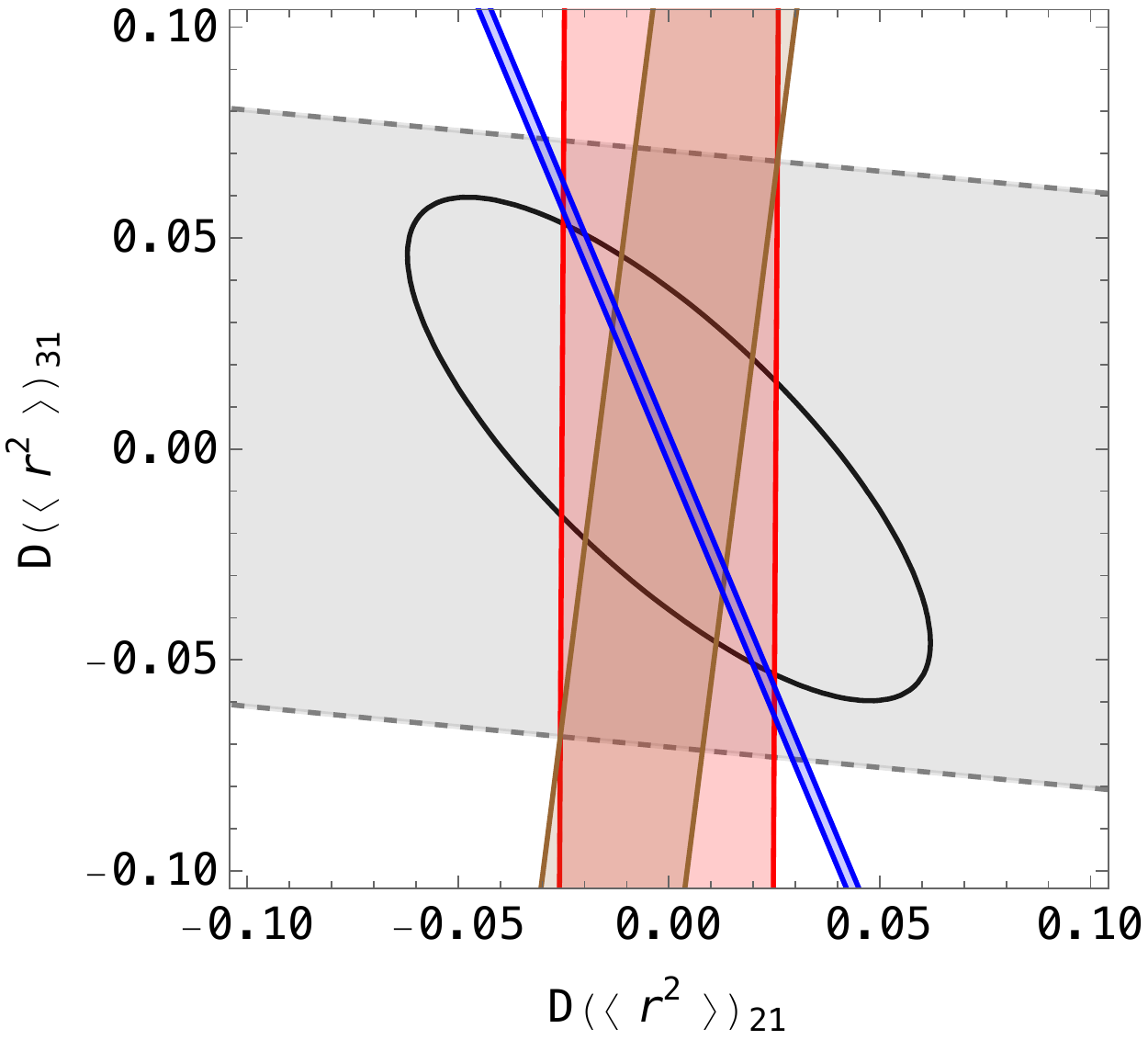}
\caption{
 The constraints on the ratios of the isotope dependences.
 (Left) The constraints on the ratios of the inverse mass differences. 
 The ellipse is given by the direct mass measurements~\cite{Huang:2021nwk,Wang:2021xhn} and the red region is given by the 2D dual King relation.
 This red region is magnified by a factor of 100.
 (Right) The constraints on the ratios of the mean square charge radius differences.
 The ellipse is given by Ref.~\cite{Angeli:2013epw}.
 The grey region is constrained by the 3D dual King relation without any assumption of the additional isotope shift.
 The blue, red, and brown regions are constrained by the assumption of the second order perturbation of the field shift, the second order perturbation of the mass shift, and the particle shift, respectively.
}
\label{FigDual}
\end{figure}

The 1 $\si$ region of $\de\mu$ ratios and the constraint by the 2D dual King relation are shown on the left of Fig.~\ref{FigDual}.
We introduce the ratio of isotope dependence for Ca as $D(\et)_{ij}= \de\et_{40+2i,40+2(i-1)}/\de\et_{40+2j,40+2(j-1)}$.
The ratios $D(\mu)_{31}$ and $D(\mu)_{21}$ have the errors of about $10^{-6}$ and $10^{-7}$, respectively.
Its best fit point is about $(0.831, 0.910)$, but we shift it to $(0, 0)$.
The constraint obtained with the linear fit is the red region, where the width is magnified by a factor of 100.
Hence, the ratios are constrained on the line indicated by the linear relation.
The isotope shifts limit the ratio $D(\mu)_{41}$ at the same time.
Because it follows the linear relation with other fit parameters, the constraint is as strong as and almost independent of $D(\mu)_{31}$.
The ratios $D(\avg{r^2})_{ij}$ are similarly constrained on a line with a width of $O(10^{-7})$, while their errors are calculated as $O(10^{-2})$ with Ref.~\cite{Angeli:2013epw}.

The similar constraints to $\de\avg{r^2}$ are shown on the right of Fig.~\ref{FigDual} with the 3D dual King relation.
The best fit of $(D(\avg{r^2})_{31}, D(\avg{r^2})_{21})$ is approximately $(-0.740, 0.316)$.
We obtained the figure by adding $D(\avg{r^2})_{41}$ to $\ch^2$ because the isotope shifts are bounded on the 2D plane in 3D space.
The constraints on the figure are affected by this precision, so that they are not as strong as the constraints with the 2D relation.
In this analysis, we include also the nuclear masses in $\ch^2$.

Without specifying the origin of the unknown nuclear factor, the constraint is drawn by the gray region, which involves the ellipse of the ratios.
If we suppose particular sources as the origin of the higher order isotope shift, we can obtain stronger constraints.
If the higher order one is generated by the second order perturbation of the mass shift and the particle shift, these constraints cover a similar region, which are shown by the red and the brown region, respectively.
Here, their best fit points are moved to the origin, namely, the best fit point of the real data.
When we evaluate the constraints with future isotope shift data, they do not have to cover similar region.

If the additional isotope shift is given by the second order perturbation of the field shift, the constraint is given by the blue region in the figure.
In this case, both $\de\avg{r^2}$ and its squares need to simultaneously satisfy the same linear relation.
Then, the allowed region given by the isotope shifts is tighter than the other assumptions.
Here, we compared the constraint given by other experiments with the one given by the isotope shifts.
If we have real data, we test the consistency between these independent constraints with the sum of their $\ch^2$.

\begin{figure}[hbt]
\centering
\includegraphics[scale=0.7,clip]{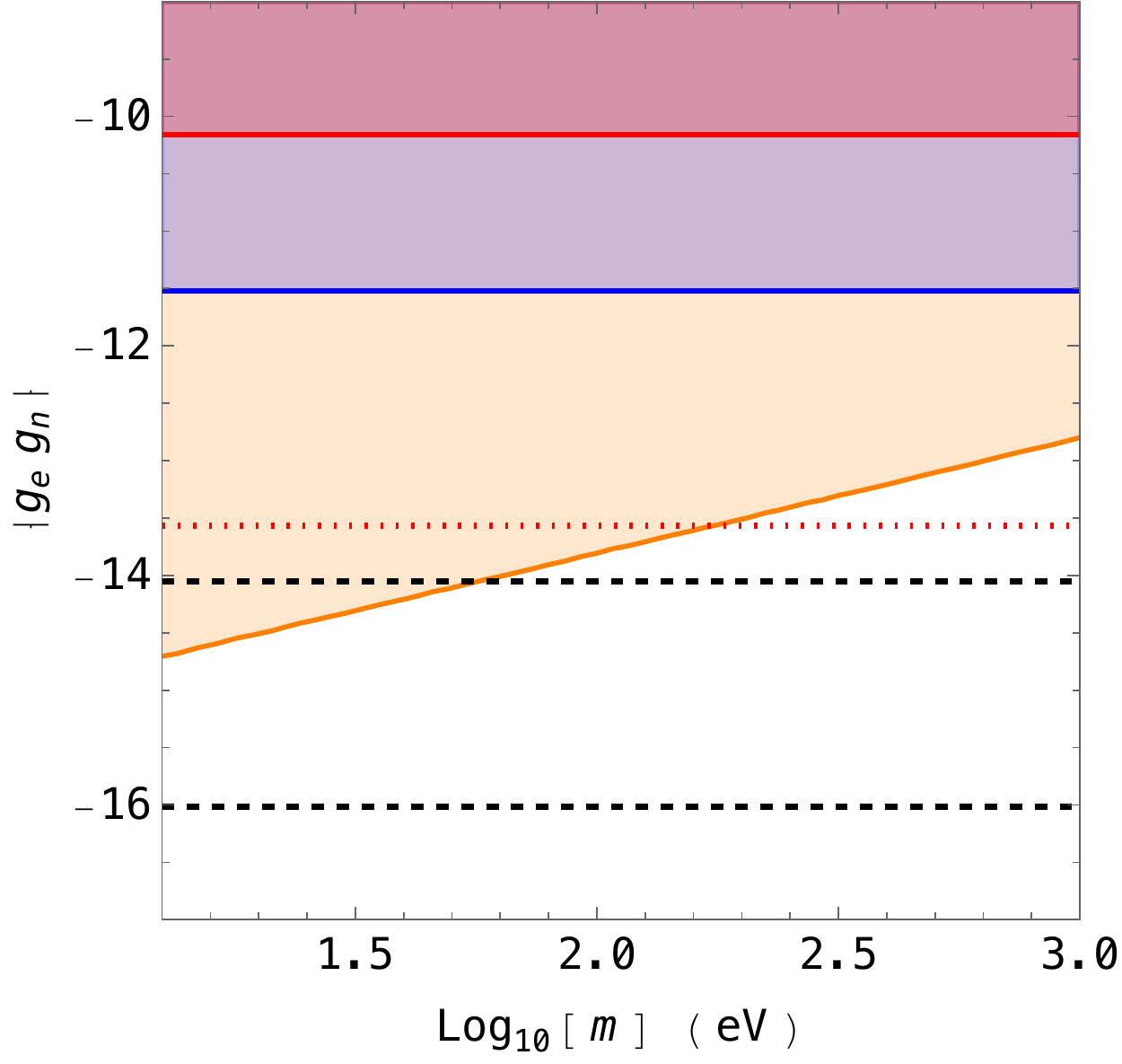}
\caption{
 The 95\% confidence level upper bounds for the light new boson coupling between electrons and neutrons.
 The red and the blue regions are respectively excluded by the isotope shift measurements of Ca and Yb in Refs.~\cite{2005.00529,2201.03578}.
 The orange region is excluded by the product of the individual constraints on the new boson couplings to electron~\cite{Morel:2020dww} and neutron~\cite{2103.05428}.
 The red dotted line is the expected reach with Ca at a precision of 10 mHz~\cite{2005.00529}.
 The upper black dashed line is the upper bound evaluated by the 2D dual King relation with the Ca transitions at the 1 Hz precision.
 The lower black dashed line is the upper bound of the 10 mHz precision with the same transitions and the 3D relation.
}
\label{FigNp}
\end{figure}

The constraints on the weakly interacting new light boson are straightforward.
Here, we explicitly subtract particle shift such as Eq.~\eqref{EqDualplus} to obtain the upper bound.
The current upper bounds of the 95\% confidence level with isotope shift measurements are provided by Refs~\cite{2005.00529,2201.03578}.
They, respectively, obtained $|g_e g_n| < 6.9\times 10^{-11}$ and $3.0\times 10^{-12}$ with Ca and Yb.
The red and the blue regions show these bounds in Fig.~\ref{FigNp}. 
They are weaker than the bound calculated by the product of individual constraints to the new force with electron~\cite{Morel:2020dww} and neutron~\cite{2103.05428}.
Here, we have assumed that the new force is mediated by a vector boson, and this combined bound is obtained by simply multiplying their 95\% upper bounds.
For the electronic factors of the particle shift, we employ the values at the new boson mass of 1 keV with FAC given by Ref.~\cite{2102.02309} as they are constant below 1 keV.

The 95\% upper bound $|g_e g_n| < 8.8\times 10^{-15}$ is obtained if the data are consistent with the 2D linear relation at 1 Hz precision, as shown by the upper black dashed line.
Since the fake data set has a small non-zero $\ch^2$, the bounds are slightly different for the sign of the coupling product.
We simply averaged the negative and the positive sign upper bounds.
The expected bound with 10 mHz, which is obtained by the 2D original King relation of Ca in Ref.~\cite{2005.00529}, is $|g_e g_n|<2.7\times 10^{-14}$.
This is indicated by the red dotted line in the figure.
Since they employed the close transitions ${}^2 S_{1/2}\to {}^2 D_{5/2}$ and ${}^2 D_{3/2}$, the non-linearities are suppressed, as we can see in Ref.~\cite{2004.11383} for Yb.
In this case, it is plausible that this bound is similar to the above 1 Hz bound here.

For the 3D relation with a precision of 10 mHz, we obtain the upper bound $|g_e g_n|<9.7\times 10^{-17}$ of the 95\% confidence level, which is drawn as the lower black dashed line in the figure.
The bound is improved by two orders of magnitude as the precision is also improved from 1 Hz to 10 mHz.
In this example, we find that the upper bound linearly improves as the precision of measurements is improved, even eliminating an additional isotope shift with the 3D relation.

\section{Conclusion}
\label{SecCon}

We introduced a new linear relation that the isotope shift follows.
This relation is derived by exchanging the transition dependence of electron and the isotope dependence of nucleus in the original King relation.
Therefore, we call it the dual King relation.

The new relation gives us similar information for the linear fit of the isotope shift as the original King relation.
In the fit, the different transitions are aligned on a linear relation in a space spanned by distinct isotope pairs.
For an additional transition in the original King relation, we need an additional linear relation and fit parameters, but any transition follows the same dual King relation.
This makes the analysis simpler than the original King relation when we measure the precision isotope shifts with many transitions in the future.

The fitting coefficients of the dual King relation consist of the nuclear isotope dependences.
The coefficients of the 2D relation are given by the mean square charge radius and the inverse mass differences.
They are measured independently by other experiments.
With precision measurements at the 1 Hz level, the ratios of the nuclear isotope dependences are constrained much more strongly than those experiments.
For the 3D relation, we can test each assumption of a higher order isotope shift with the measurements of other experiments through the fit coefficients.

We can obtain constraints on the new force coupling between electron and neutron that is similar to what we got from the original King relation.
The upper bounds of the coupling can be estimated as $|g_e g_n| < 8.8\times 10^{-15}$ and $9.7\times 10^{-17}$ with the precision of 1 Hz and 10 mHz, respectively.
They are stronger than the current strongest bound evaluated by the combination of other experiments above 100 eV and 10 eV of the mass region.

The measurements of the precision isotope shifts with the elements that have more spin-0 nuclei, like Sn, Xe, etc., are important to improve the constraints on the new boson beyond the higher order Standard Model contributions.
The dual King relation can be used to test the origin of these contributions by imposing independent constraints on nuclear isotope dependences measured in other experiments.
Furthermore, we can utilize the dual King relation to directly extract the nuclear isotope dependences of unstable nuclei.

\section*{Acknowledgments}
The author thanks M.~Tanaka for a comment on the manuscript.
This work was supported by the National Science and Technology Council, the Ministry of Education (Higher Education Sprout Project NTU-111L104022), the National Center for Theoretical Sciences of Taiwan, and the visitor program of Yukawa Institute for Theoretical Physics, Kyoto University.

%

\end{document}